**Title**: Low-carbon Lithium Extraction Makes Deep Geothermal Plants Cost-competitive in Energy Systems

**Authors:** Jann Michael Weinand,[1] Ganga Vandenberg,[1] Stanley Risch,[1,2] Johannes Behrens,[1,2] Noah Pflugradt,[1] Jochen Linßen,[1] and Detlef Stolten[1,2]

**Institutions:**

[1]Forschungszentrum Jülich GmbH, Institute of Energy and Climate Research – Techno-economic Systems Analysis (IEK-3), 52425 Jülich, Germany
[2]RWTH Aachen University, Chair for Fuel Cells, Faculty of Mechanical Engineering, 52062 Aachen, Germany

**Corresponding Author:** Jann Michael Weinand, j.weinand@fz-juelich.de, +49 175 4985402

**Conflict of Interest Statement:** The authors have no conflicts of interest to disclose.

**Abstract:** Lithium is a critical material for the energy transition, but conventional procurement methods have significant environmental impacts. In this study, we utilize regional energy system optimizations to investigate the techno-economic potential of the low-carbon alternative of direct lithium extraction in deep geothermal plants. We show that geothermal plants will become cost-competitive in conjunction with lithium extraction, even under unfavorable conditions and partially displace photovoltaics, wind power, and storage from energy systems. Our analysis indicates that if 10% of municipalities in the Upper Rhine Graben area in Germany constructed deep geothermal plants, they could provide enough lithium to produce about 1.2 million electric vehicle battery packs per year, equivalent to 70% of today`s annual electric vehicle registrations in the European Union. This approach could offer significant environmental benefits and has high potential for mass application also in other countries, such as the United States, United Kingdom, France, and Italy, highlighting the importance of further research and development of this technology.

## 1. Introduction

Lithium is crucial for the transition to greenhouse gas neutral energy systems. In 2019, over 60% of lithium produced was utilized for the manufacturing of lithium-ion batteries, the compact and high-density energy storage devices for low-carbon-emission electric vehicles, and secondarily as a storage medium for renewable energy sources like solar and wind [1,2]. In 2 °C compatible scenarios, today`s global lithium demand would be expected to grow by another 500% by 2050 [3].

However, established lithium extraction procedures like salar brine and hardrock mining are highly carbon-intensive and contribute to air and water pollution, require large amounts of water and land, and are associated with human rights violations and poor worker protection [4]. With roughly 90% of lithium extraction taking place in Australia, Chile and China, and almost 100% of its processing occurring in China, Chile and Argentina, most other countries in the world are completely dependent on lithium imports [5]. Increased production and diversification of lithium supply are needed to meet anticipated demand and improve mineral security, whereas sustainable extraction methods are required to reduce carbon intensity and environmental impacts [6,7].

One promising sustainable extraction option that involves reduced water and land footprints is hybrid geothermal plants, which combine deep geothermal power and heat production with low-carbon direct lithium extraction (DLE) [8]. Currently, pilot projects utilizing this technology are being developed in the Upper Rhine Graben (URG) in France [9] and Germany [10], Cornwall in the United Kingdom [8], and the Salton Sea in California, United States [11]. From an economic perspective, deep geothermal energy is not yet viable in low- to intermediate-enthalpy regions such as Germany [12,13] and thus not competitive with other renewable energy sources. In contrast to low-cost photovoltaics [14] and wind energy [15,16], future cost reductions are expected to be fairly low [17]. In view of the wide range of possible applications and rising lithium prices [5,18,19], dispatchable deep geothermal systems could yet play a major role in future energy systems. To assess this requires integrated energy system analysis that involves geothermal plants together with DLE.

A few studies have incorporated deep geothermal systems into decentralized energy systems, such as district heating applications [20,21] or simultaneous power and heat supply as part of integrated energy system optimization in municipalities [12,17,22], but never in combination with DLE. For lithium extraction from geothermal brines, previous studies have focused on technical and economic potential, especially for the Salton Sea Known Geothermal Resource Area (KGRA) [23–25], supply chain impacts of increased lithium supply [26], geochemical characteristics of geothermal brines in Germany (Molasse Basin [27], Upper Rhine Graben [28,29] and North German Basin [30–32]), and environmental impacts of lithium extraction [33,34].

In this article, we investigate for the first time the techno-economic impacts of installing and operating deep geothermal systems with lithium extraction in decentralized energy systems. For this purpose, we focus on the Upper Rhine Graben in Germany, whose brine lithium deposits are comparable to currently exploited evaporative brine and hard rock mining lithium operations [29,32,35,36]. An integrated energy system model, based on the open-source framework ETHOS.FINE [37], is extended to include hybrid geothermal plants (Section 2) and applied to optimize greenhouse gas-neutral energy systems of municipalities located in the Upper Rhine Graben in Germany for the year 2045 from a macroeconomic perspective (Section 3). Thus, based on expert evaluations of the key parameters of lithium extraction plants and through distinctive sensitivity analyses, we show the conditions under which deep geothermal energy with DLE will become an indispensable component of future energy systems. In Section 4, we discuss our findings in the context of the global energy transformation and derive conclusions.

## 2. Methods

In the methodology section, we first describe the energy system optimization framework used, on which the regional model for individual municipalities is based (Section 2.1). Subsequently, we address the key equations used to represent the geothermal plant (Section 2.2), as well as how hydrothermal temperatures and drilling are incorporated in the model (Section 2.3). The implementation of the DLE plant is shown in Section 2.4 along with key cost assumptions. Finally, in Section 2.5, we describe the studied municipalities from our case studies.

### 2.1. ETHOS.FNE optimization

This study utilizes a municipal energy system optimization model, which is based on the open-source *Framework for Integrated Energy System Assessment* (ETHOS.FINE) Python package [37]. The model provides a framework for modeling, optimizing, and assessing regional energy systems using high-resolution generation and consumption data. The objective of the model is the minimization of total annual costs (TAC) for supplying all demand sectors of a municipality while considering the technical and environmental constraints for a greenhouse gas-neutral renewable energy system in 2045. The costs are composed of the total annual costs of all built renewable power generation technologies, conversion technologies, and storage technologies, as well as sources/sinks (e.g., photovoltaic panels or lithium demand), and are determined using each technology's per unit capital costs, annuity factor, number of built installations, and operation and maintenance costs. The total costs of components may be negative, as revenues from sources/sinks are included in the operational costs (e.g., through electricity or lithium carbonate sales). The optimization is performed from the perspective of a central planner with perfect foresight. Although the model can also be used for analyses at the NUTS-3 administrative level or higher, those presented in this work take place at the municipal level. The application of a hierarchical clustering approach with the *Time Series Aggregation Module* (TSAM) [38] with 60 periods and 16 segments enables the analysis of a high number of energy systems at an hourly resolution (8760 h) without significant accuracy losses (mean deviation in optimized total annual costs: 0.3%).

The optimization model includes onshore wind, rooftop photovoltaics (PV), open-field photovoltaics (OFPV), biomass, biogas, and waste, and is extended by deep geothermal plants and the commodities of lithium and lithium carbonate ($Li_2CO_3$) (see Figure 1). Regional potentials for rooftop and open-field PV, as well as wind, are determined using the *Tool for Regional Renewable Potentials* (TREP) [39]. Energy demand sinks are households, the trade commerce and service sector (TCS), and industry, as well as their respective commodities. Industrial energy demand consists of the demand for electricity, heat, and process heat. Process heat is implemented in three different forms: low-temperature for up to 100 °C, medium-temperature for between 100 and 500 °C, and high-temperature for processes above 500 °C. For the regional demand time series, top-down demand data [40] is regionalized based on employment, population, and $CO_2$ emissions data.

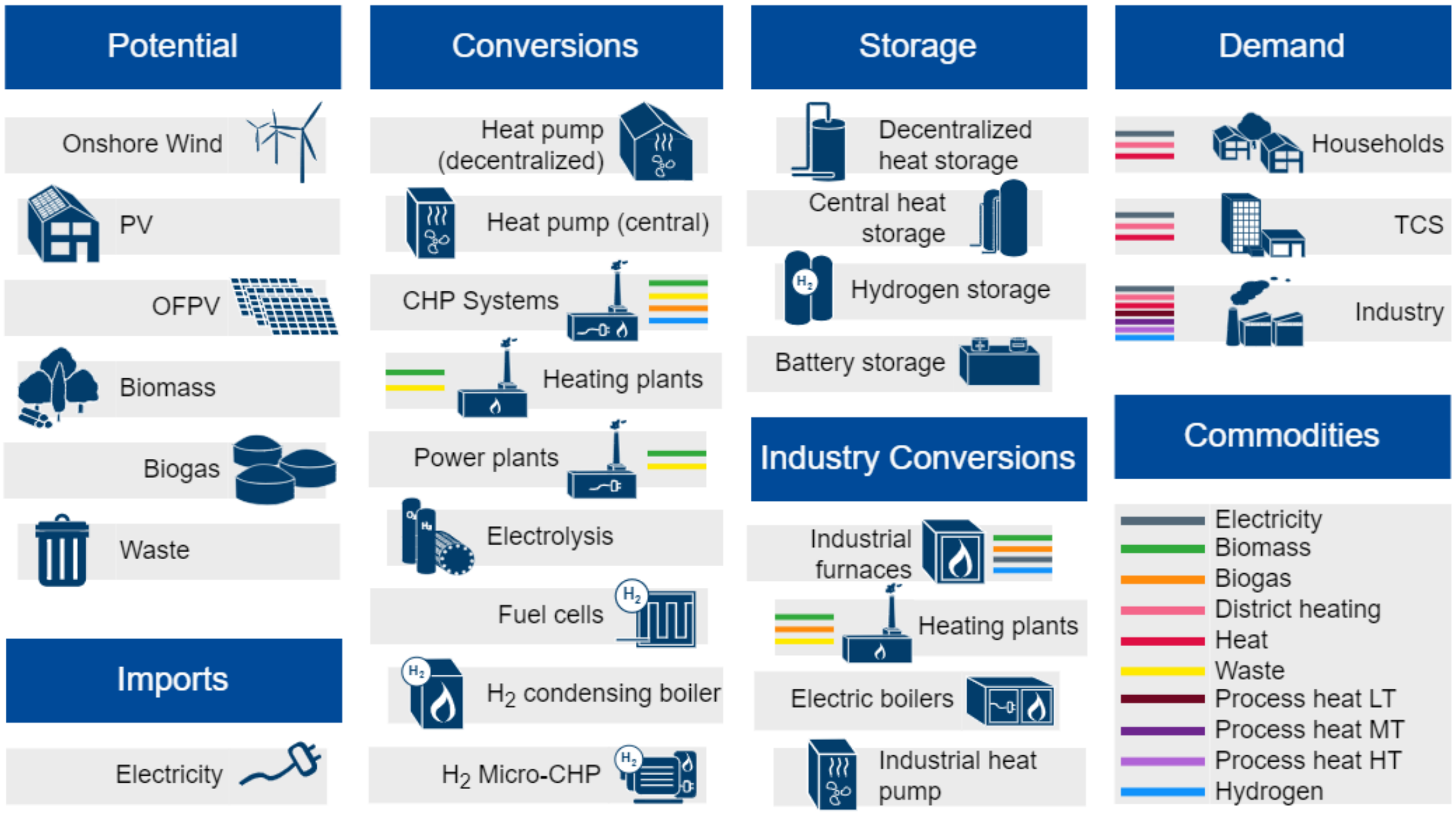

*Figure 1: Components of the energy system optimization model. These include renewable potentials, imports, conversion, and storage technologies, as well as demand sectors. Commodities that are in demand or supplied are indicated with different colors, and only if the technology involves more than one.*

## 2.2. Deep geothermal plant model

A geothermal plant utilizes thermal energy in deep hydrothermal aquifers to produce heat and/or electricity (see Figure 2). The power generation $P_{el}$ of the Organic Rankine Cycle plant and the heat generation $\dot{Q}_{th}$ of the district heating plant per time step t are determined as follows [12].

$$\dot{V}_B \cdot \rho_w \cdot c_{p,w} \cdot \big(T_{PW}(t) - T_{ORC}(t)\big) \cdot \eta_{el} = P_{el}(t) \qquad \forall t \qquad (1)$$

$$\dot{V}_B \cdot \rho_w \cdot c_{p,w} \cdot \big(T_{ORC}(t) - T_{DHP}(t)\big) \cdot \eta_{th} = \dot{Q}_{th}(t) \qquad \forall t \qquad (2)$$

where $\dot{V}_B$ is the volumetric flow rate of the geothermal brine in l/s, $\rho_w$ the mean density of the geothermal water in kg/l, $c_{p,w}$ the mean heat capacity of the geothermal water in kJ/(kg·K), and $T_{PW}$, $T_{ORC}$, and $T_{DHP}$ the temperatures in the production well and after heat transfer to the Organic Rankine Cycle or the district heating network, respectively. As the flowrate $\dot{V}_B$ can vary greatly depending on local geological conditions, the mean flow rate of 75 l/s for existing deep geothermal systems in Germany is utilized in this model unless stated otherwise (see scenarios). A mean heat density $\rho_w$ of 0.95 kg/l and mean heat capacity $c_{p,w}$ of 4.31 kJ/(kg·K) are assumed [12]. The minimum injection temperature is 50 °C, which directly affects the temperature after heat transfer to the district heating network $T_{DHP}$. The optimization model can choose to build a district heating plant and/or Organic Rankine Cycle plant and decides how to allocate the heat source between the two if both are built. The efficiency of the ORC plant $\eta_{el}$ is assumed to be 10%, with 65% assumed for the efficiency $\eta_{th}$ of the district heating plant [12].

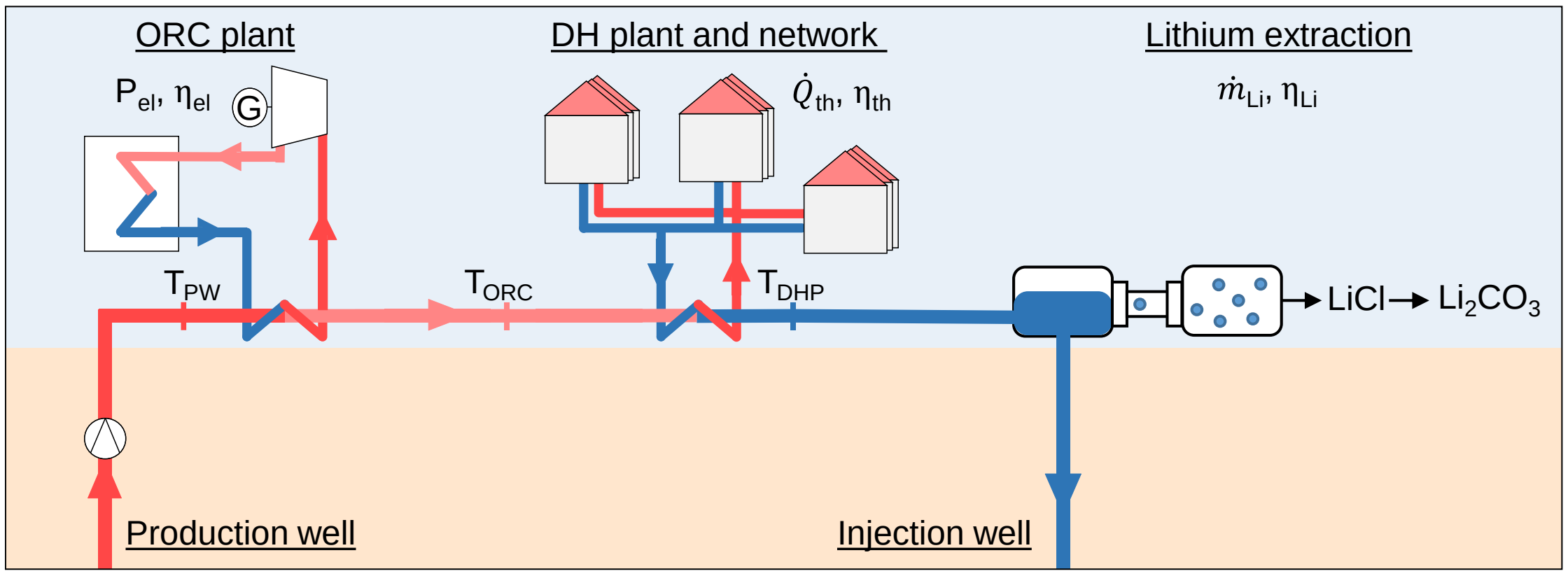

*Figure 2: Schematic representation of the Organic Rankine Cycle (ORC) and district heating (DH) plant and network, as well as lithium extraction considered in this study. Geothermal brine is pumped up by the production well pump and fed to a heat exchanger, where it heats a working fluid in the ORC plant, which in turn drives an electric generator, producing electricity. The brine then goes through another heat exchanger at the district heating plant, which supplies heat to the district heating network. The cooled brine can then be transported to the lithium extraction plant and brought in contact with a lithium-selective adsorbent that binds with the lithium ions. The lithium is then separated from the adsorbent and can be upgraded to lithium carbonate, after which the cooled lithium-depleted brine is returned underground via the injection well.*

## 2.3. Hydrothermal temperatures and drilling

Drilling costs account for the majority of geothermal plant investment costs, with a share of up to 70% [12]. As these cost functions are non-linear (see Eq. 3 [12]) the optimization model must select one drilling depth from amongst a set of up to 400 discrete options in steps of 10 m from 1000 m, and up to 5000 m. The lower limit of 1000 m is used, as lithium reserves are only present at greater depths. It is assumed that economies of scale apply to these drilling costs, with the cost of the second well being 90% those of the first. The drilling costs are calculated using the drilling depth $z_D$ in meters, as well as the distance between the production well and injection well $d_D$ in meters:

$$C_D = 610{,}000€ + 1.015 \cdot 1.198\, e^{0.00047894 \cdot \sqrt{z_D^2 + d_D^2}} \cdot 10^6€ \tag{3}$$

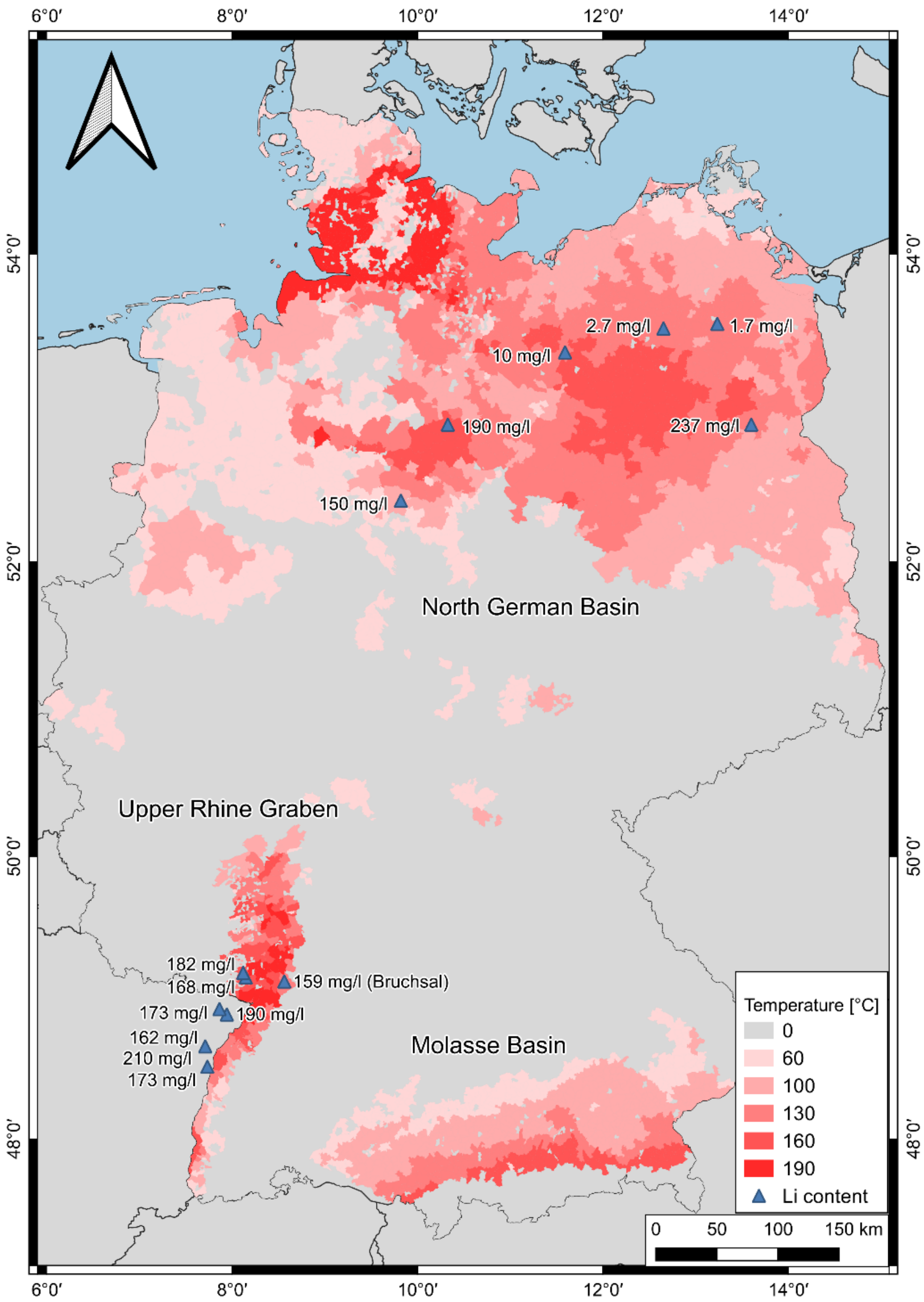

*Figure 3: Achievable hydrothermal temperatures in Germany at a depth of up to 5,000 m [41] and measured lithium contents [28,31,32,42,43].*

The selected drilling depths then dictate the maximum achievable hydrothermal temperature in the optimization, the theoretical maxima of which can be found for German municipalities [13] up to a depth of 5000 m in Figure 3. The assumed mean temperature gradients for the major geothermal basins, the Molasse Basin, the North German Basin, and Upper Rhine Graben, are 32  C/km, 35  C/km, and 43 °C/km, respectively. Locally, however, the temperature gradient for the Upper Rhine Graben may be much higher [44], particularly at depths of up to 3 kilometers, with average values of up to 110  C/km. Therefore, for the Upper Rhine Graben the assumed average temperature gradient has been divided into three sections with 47 °C/km until a depth of 1900 m, 41 °C/km between 1900 m and 3250 m, and 33 °C/km from 3250 m and above.

### 2.4. Direct lithium extraction

After the heat exchange with the Organic Rankine Cycle and district heating network, the cooled brine is transported to the lithium extraction plant and brought in contact with a lithium-selective adsorbent that binds with the lithium ions. The lithium is then separated from the adsorbent and upgraded to lithium carbonate, and the cooled lithium-depleted brine is returned underground via the injection well (Figure 2). In the optimizations, a mean lithium concentration of 175 mg/l is assumed based on measured data for the Upper Rhine Graben (Figure 3). The quantity of lithium extracted from lithium-bearing geothermal brines is determined using Eq. 4:

$$\dot{V}_B \cdot C_{Li} \cdot \eta_{Li} = \dot{m}_{Li} \qquad (4)$$

where $\dot{V}_B$ is the brine flow rate measured in l/s, $C_{Li}$ the concentration of lithium in the brine measured in mg/l, $\eta_{Li}$ the extraction efficiency, with the final product being elemental lithium $\dot{m}_{Li}$ measured in mg/s. After the extraction, the lithium is processed with a conversion factor of 5.324 [23] into lithium carbonate, which is a largely traded raw material to produce, e.g., lithium-ion batteries [6].

Economic and technical data on lithium extraction from geothermal brines is scarce and therefore subject to major uncertainties. Whilst we were able to find literature values for all needed parameters, we assessed the impact of each of these in extensive sensitivity analyses (see main text). Furthermore, we assume fixed contract prices for the lithium carbonate market prices of between 8500 €/t and 25,500 €/t. The average annual lithium carbonate price for fixed contracts has more than doubled since 2020, reaching 17,000 €/t in 2021 [5]. Typically, such fixed contracts for lithium carbonate last three to five years [4]. More recently, spot prices have shown even greater volatility, rising from roughly 5500 €/t lithium carbonate in September 2020 to over 76,000 €/t in September of 2022 [5]. However, spot prices are typically higher than contract prices, and studies anticipate that in the long-term, the market price will be significantly lower than the current spot market price [45,46]. Lithium carbonate market volatility has been observed in the past, with fixed contract prices increasing from 2015 to 2018 and then decreasing sharply until 2020. The 2015 and present spikes in pricing can be attributed to “unexpected and explosive EV market growth” [46], while the latter is also attributable to the COVID-19 pandemic. Future market prices will largely be determined by available reserves, as well as the growth of electric vehicle sales. In the long-term, lithium carbonate pricing could decrease to as low as 10,000 €/t [46].

### 2.5. Case studies

A total of 330 municipalities in the Upper Rhine Graben in Germany (see Figure 3) have achievable hydrothermal temperatures of 60 °C or more. We investigate the optimal energy systems of these municipalities with and without the DLE option in the Mean URG scenario (see Table 1 and Section 3.3). The municipalities have about 4.5 million inhabitants, and the mean population density is 400 inhabitants/km². The three most populous cities in the Upper Rhine Graben are Karlsruhe (about 312,000 inhabitants and 1,800 inhabitants/km²), Mannheim (310,000 inhabitants and 2,140 inhabitants/km²), and Freiburg im Breisgau (230,000 inhabitants and 1,510 inhabitants/km²).

*Table 1. Energy system optimization scenarios considered in this article. The baseline scenario contains proven existing values of the Bruchsal location, as well as the mean or most probable values for the direct lithium extraction (DLE) plant based on literature and expert opinions. Worst and best case scenarios include the worst or best values from the literature or existing plants, respectively. The optimistic scenario represents a state that might be reached and applies mean values between the baseline and best case scenarios. The mean URG scenario is applied to the energy system optimizations of all municipalities of the Upper Rhine Graben in Section 3.3 and represents the mean values of all existing plants in the region, as well as the DLE values from the Baseline scenario. However, as the data on the maximum achievable temperatures are available for each municipality, the temperature is specific to each of these and ranges from 60 to 190 °C in this scenario.*

| Parameter | Worst case scenario | Baseline scenario | Optimistic scenario | Best case scenario | Mean URG scenario |
|---|---|---|---|---|---|
| Flow rate [l/s] [47] | 24 | 24 | 82 | 140 | 75 |
| Maximum wellhead temperature [°C] [47] | 65 | 131 | 176 | 220 | 60–190 |
| Lithium concentration [mg/l] [28,29,31,42,43,48] | 86 | 159 | 198 | 237 | 175 |
| DLE CAPEX [M€] [49] | 31.2 | 20.8 | 15.8 | 10.9 | 20.8 |
| DLE OPEX [€/t] [23] | 8,000 | 4,000 | 3,000 | 2,000 | 4,000 |
| DLE efficiency [-] [50] | 50% | 70% | 80% | 90% | 70% |
| Li carbonate market price [€/t] [5] | 8,500 | 17,000 | 21,250 | 25,500 | 17,000 |

The optimal energy system of one specific municipality in the Upper Rhine Graben, Bruchsal, is examined in four other scenarios (Table 1, Section 3.1) and various sensitivity analyses (Section 3.2). With a population of about 44,800 inhabitants and a municipality area of 93 $km^2$, the population density of Bruchsal is roughly 480 inhabitants/$km^2$ [51]. The maximum renewable potentials for the municipality include 75 $MW_{el}$ of onshore wind, 31 $MW_{el}$ of open-field PV, and 290 $MW_{el}$ of rooftop PV [39]. In addition to these potentials, Bruchsal has an already installed capacity of 1.22 $MW_{el}$ for open-field PV and 24 $MW_{el}$ for rooftop PV, which were included in this study as existing capacity. The total electricity demand of the municipality is approximately 480 $GWh_{el}$ and the total heating demand is roughly 465 $GWh_{th}$. Residential heating demand comprises roughly 45% of the total heat demand, while industry electricity demand makes up the largest portion of the total electricity demand at about 31%. The stated total electricity demand also includes the optimization results of ca. 183 $GWh_{el}$ for storage losses and electricity conversion to heat, process heat, and hydrogen ($H_2$).

# 3. Results

## 3.1. Direct lithium extraction benefits deep geothermal plants

The Bruchsal geothermal well in the Upper Rhine Graben is currently being investigated in pilot projects to identify qualified lithium-selective adsorbents, determine reservoir sustainability, assess environmental impacts, and evaluate whether lithium extraction from geothermal brines can be economically competitive with lithium sourced from South America and Australia using conventional methods. Bruchsal has a favorable lithium content (159 mg/l), temperature gradient (on average 43 °C/km), and reservoir temperature (131 °C) for such a project [29,41] and is therefore investigated here as a first case study in four scenarios (Table 1). Further information on the demand and supply structure of the municipality can be found in the Methods section.

Deep geothermal plants for power and heat generation alone are only cost-competitive under very favorable conditions and thus are not installed in optimal energy systems due to the low achievable flow rate in Bruchsal. This finding is in line with previous analyses using different energy system optimization models [12,17]. If no geothermal plant is built, most of the electricity or heat will be provided by onshore wind, rooftop and open field photovoltaic, or heat pumps, respectively; see the worst case scenario in Figure 4, which results in the same energy system as the baseline scenario without DLE.

However, depending on the geological characteristics of the geothermal source, the option of lithium extraction and sale makes deep geothermal plants cost-competitive (see the baseline, optimistic, and best case scenarios in Figure 4). Although the deployment of geothermal plants increases the annual costs for the energy supply technologies, this is offset by the revenues from lithium carbonate sales in quantities of 450–3,900 tons, even leading to negative annual costs in the best case scenario. Depending on the flow rate, wellhead temperature, and lithium concentration and extraction efficiencies, the geothermal plant displaces 6–29% of rooftop photovoltaics. The district heating plant is favored over the Organic Rankine Cycle, leading to the latter only being built in the best case scenario, assuming excellent hydrothermal resources. Due to the base load capacity of the geothermal plant and the large district heating displacing 7–75% of the heat pumps, overall power generation and the need for heat and electricity storage decreases.

The developed model of the geothermal plant reflects the reality fairly well. If the real temperature of the Bruchsal plant (123 °C) is fixed in the baseline scenario, similar values are chosen by the model with a 2470 m drilling depth compared to a 2542 m one in reality (-3%), as well as 4.66 $MW_{th}$ district heating plant capacity compared to 5.7 $MW_{th}$ (-18%) [12]. In this assessment, it is important to keep in mind that average parameters were assumed to ensure the applicability of the developed model for every municipality in Germany, e.g., for temperature gradients and efficiencies, etc. However, the most uncertain aspect of a geothermal project, the drilling costs, cannot be estimated very accurately using our model. Here, the model results of 11.4 M€ are 41% higher than the real costs of 8.1 M€ [12]. For the costs, a safe conservative estimate had to be made in our model, as geothermal projects can become more expensive than initially estimated due to unexpected costs arising. This means that the valuation of geothermal plants in this study could be slightly underestimated for specific regions.

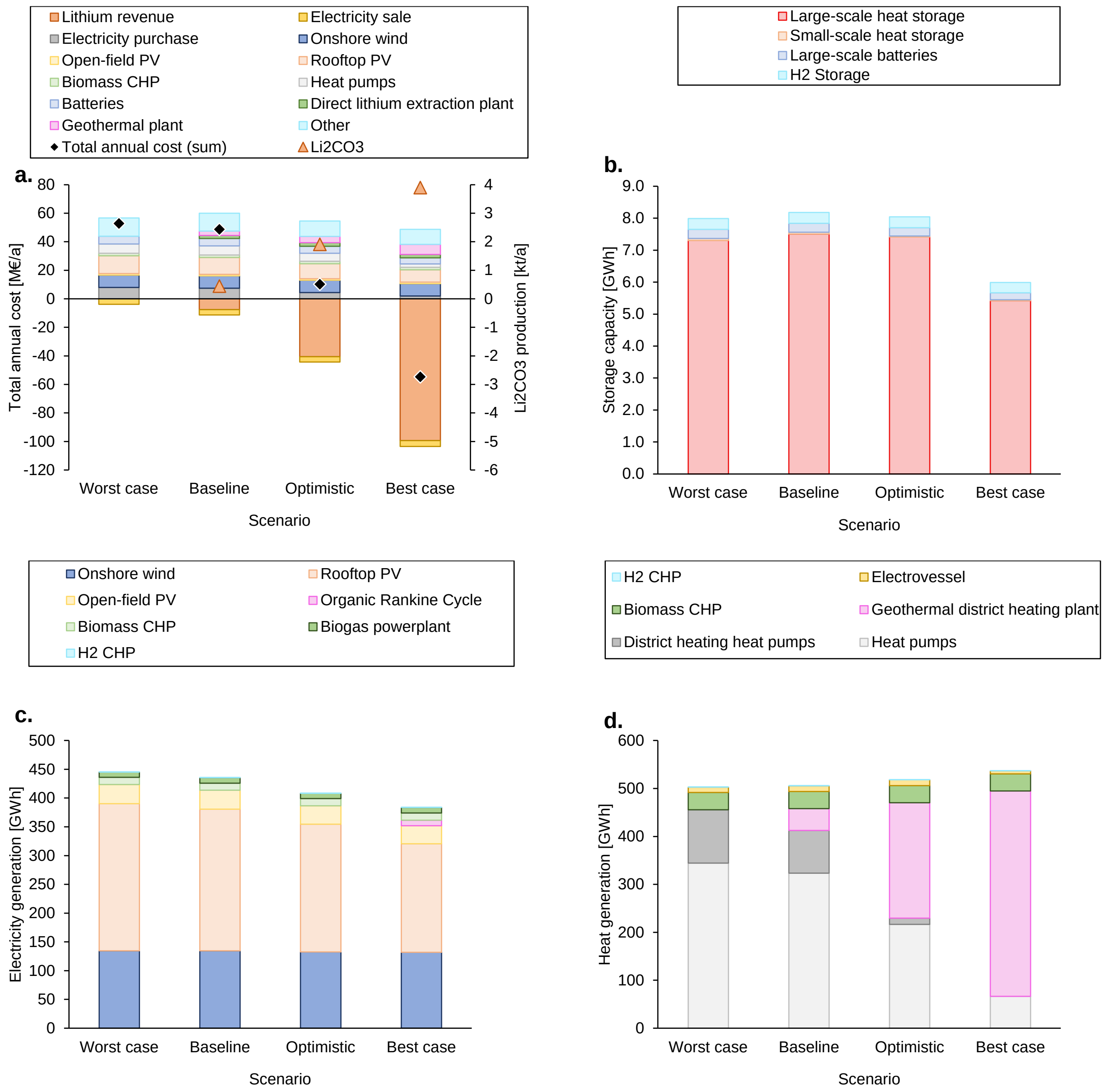

*Figure 4: Optimized energy system by 2045 in the worst case, baseline, optimistic, and best case scenarios for the municipality of Bruchsal. The different panels show the total annual cost (a), storage capacities (b), electricity generation (c), and heat generation (d) for the cost-optimal energy systems in the different scenarios.*

## 3.2. Cost-competitiveness even under pessimistic conditions

So far, the focus has been on the characteristics of the existing plant in Bruchsal. As this site has the lowest flow rate of all existing plants in Germany and thus tends to underestimate the potential of deep geothermal energy in the baseline or worst case scenarios, we now consider the mean values of geothermal plants in the Upper Rhine Graben for several sensitivity analyses of the Bruchsal energy system (wellhead temperature of 115 °C; for the other parameters, see the mean URG scenario in Table 1). Unlike the worst case scenario, in which the combination of unfavorable parameter values resulted in no geothermal system being installed despite the DLE possibility, in the sensitivity analyses we change only one parameter at a time to understand the individual effects on energy system design and costs.

Geothermal plants with lithium extraction remain competitive in energy systems if only individual parameter values are varied and otherwise average values assumed. As geothermal energy and lithium procurement are directly correlated with the flow rate, changes in this assumption significantly impact the results (Figure 5). The mean flow rate of geothermal plants in Germany of 75 l/s differs greatly depending upon local geological conditions. In Germany, brine flow rates range from 24 l/s at the Bruchsal geothermal plant to up to 150 l/s in the Molasse Basin [12]. Increasing flow rates is achievable through additional drilling as the operator of DLE pilot plants Vulcan Energy Resources Ltd. intends to achieve flow rates of 100–120 l/s in the Upper Rhine Graben [49]. However, as indicated by the results of the sensitivity analysis, even at greatly reduced flow rates, combined geothermal–lithium plants are still beneficial in a cost-optimized energy system.

Another significant assumption is the utilization of the mean lithium concentration in geothermal brines measured by previous studies, given the lack of publicly-available data. However, this neglects the fact that measured lithium contents in geothermal brines vary greatly by location. This may have an especially high impact on the results for the North German Basin, as the lithium deposits in that area are highly concentrated and the measured contents range from 0 to 237 mg/l [31] which is another reason why we chose the Upper Rhine Graben for our investigation. Furthermore, although experts assume lithium concentrations of 0 mg/l in the Molasse Basin, further research may also reveal lithium deposits there.

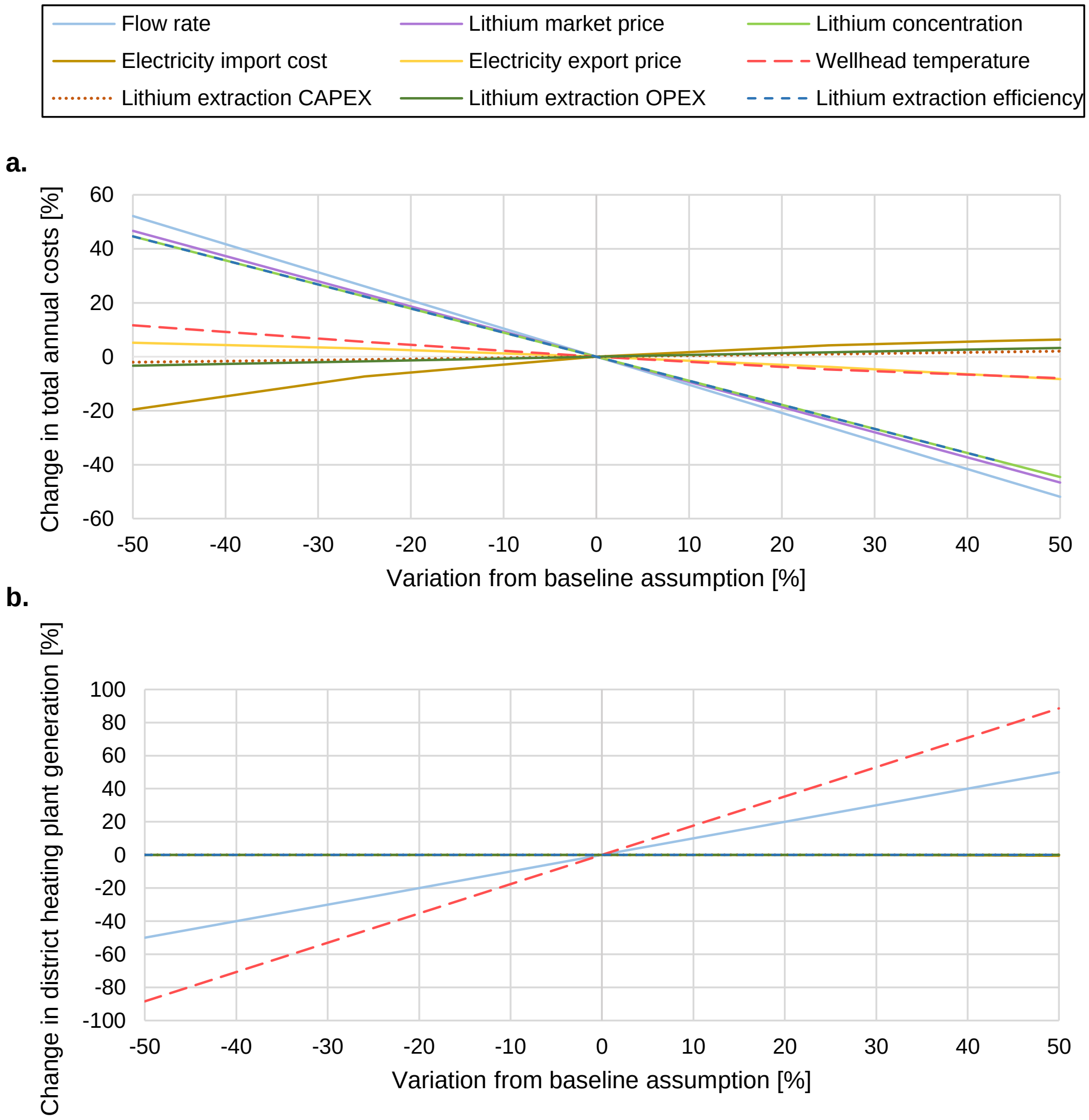

*Figure 5: Impacts of parameter variations on the design of the optimal energy system in Bruchsal. Panel (a) shows the effect of the sensitivity analyses on the total annual costs and panel (b) on the district heating plant generation. As the Organic Rankine Cycle was not installed in these analyses, its generation is not shown in the figure. The largest impact on costs comes from parameters that directly influence lithium carbonate production and sales, such as the flow rate, as well as the lithium extraction efficiency, market price, or concentration. The impact of the flow rate is the largest, as it also directly affects the maximum achievable district heating plant capacity and generation. Apart from the flow rate and wellhead temperature, the other analyzed parameters have no significant influence on the district heating plant design.*

When conducting this study, many questions also arose surrounding the economics of DLE, its efficiency, and the market price of lithium carbonate. The extraction efficiency rates in the literature vary from 50–90% [10,49] and have a significant impact on total system costs (Figure 5). The same applies to the market price of lithium carbonate, which has increased substantially in recent months. The U.S. Geological Survey estimates an average annual lithium carbonate price of 17,000 €/t for fixed contracts in 2021, which is more than double the same value in 2020 [5].

However, the spot market price for September 2022 was up to roughly 76,000 €/t and is forecast to increase [52].

Sustainable low-carbon lithium may also command a premium price compared to lithium from conventional extraction due to growing demand for low-carbon products. This demand is present in the automotive sector, with a push for electric vehicle manufacturers to decarbonize supply chains, including Volkswagen and Toyota, which have set the lofty goal of eliminating carbon emissions from their value chains [53]. The commercial interest in low-carbon lithium has already been proven in the form of offtake agreements for geothermal lithium signed by Renault, Volkswagen, Umicore, LG Energy Solutions, and Stellantis [54]. As the lithium market price has a significant impact on overall costs, such premium pricing could further improve the economics of energy systems, including combined geothermal–lithium plants.

The operating expenses (OPEX) and capital expenses (CAPEX) of DLE plants have a negligible effect on the energy system design and costs. The operating expenses identified during the literature review vary from just under 2000 €/t, per Vulcan Energy [49], to roughly 4000 €/t, as reported by the US Department of Energy [23], to up to roughly 8000 €/t per a discussion with experts. CAPEX are also quite uncertain: although we utilized a value of 20,800 M€, the actual CAPEX value for such a project could significantly differ.

### 3.3. Large-scale impacts of geothermal plants with lithium extraction

In contrast to the previous sensitivity analyses, we now optimize the energy systems of all 330 municipalities of the Upper Rhine Graben in the Mean URG scenario. This scenario utilizes the actual maximum wellhead temperature specific to each municipality, rather than being fixed at 115 °C. Even without the option of building a DLE plant, deep geothermal systems were developed in 152 of 330 municipalities (46%). These municipalities have medium- to high-enthalpy resources with a hydrothermal temperature range of 130–190 °C and an average temperature of 131 °C. This result is in line with the findings of previous studies [17] and demonstrates that for sites with very suitable conditions, deep geothermal plants are cost-competitive with conventional energy sources. All 152 municipalities installed district heating systems, whereas Organic Rankine Cycle plants were built in 113 of 330 municipalities (34%).

With the option of building a lithium extraction plant and the added revenue from the sale of lithium carbonate, deep geothermal plants are cost-competitive in all 330 municipalities. On average, the total annual costs are reduced by 22.4 M€/a or 1000% for a municipality in the URG, illustrating the added benefit of combined geothermal-lithium plants (see Figure 6, “> 100%” cost decrease means that municipalities make profit). Especially in smaller communities, the profit from the lithium sales obviously has a particularly strong effect. Key electricity generation technologies of the 330 municipalities include rooftop and open field photovoltaics (with average capacities per municipality of roughly 67 MW and 23 MW for each type, respectively) and onshore wind turbines (average capacity: 11 MW), and to a lesser extent deep geothermal Organic Rankine Cycle plants (average capacity: 0.9 MW), whereas heat is primarily supplied by heat pumps and deep geothermal district heating plants.

The development of Organic Rankine Cycle plants is associated with municipalities that have low or no onshore wind and PV potential and high achievable hydrothermal temperatures, whereas district heating plants are more favorable in larger and more densely-populated municipalities like Karlsruhe, Heidelberg or Mannheim. Compared to the optimal systems in the scenario without DLE, deep geothermal systems primarily displace rooftop PV capacity (average of 2.0 MW or -24% of original capacity), followed by open-field PV (1.9 MW or -14%) and onshore wind (1.2 MW or -29%), whereas district heating plants primarily displace heat pumps (2.1 MW or -64%). The tendency to displace more photovoltaics, even though the cost of electricity generation is lower, can be explained by the higher system integration costs compared to wind power [55].

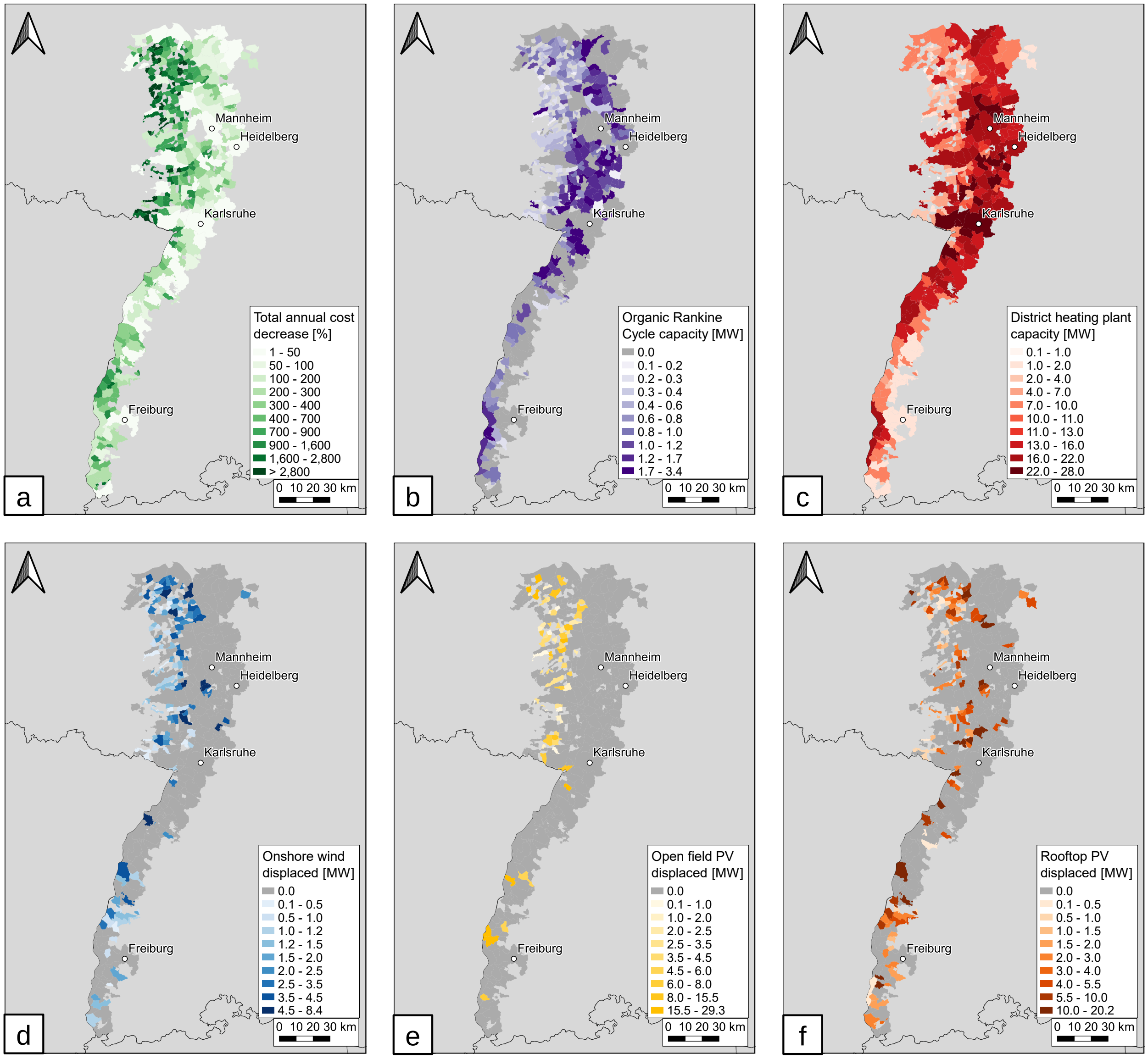

*Figure 6: Cost-optimal energy systems of 330 municipalities in the Upper Rhine Graben with the option of direct lithium extraction compared to energy systems without this option. The figure panels show how the total annual cost (a), capacities of Organic Rankine Cycle (b), district heating plant (c), onshore wind turbines (d), open field (e), and rooftop (f) photovoltaics are affected if the option to install direct lithium extraction is given compared to optimal energy systems without this option.*

If every municipality in the URG were to install a hybrid geothermal plant with lithium extraction, ca. 510 kt of lithium carbonate could be produced, which lies well within the range of current estimates. With a typical electric vehicle lithium-ion battery pack (NMC523 type) containing ca. 8 kg of lithium [56] enough to manufacture over 11.9 million battery packs annually, greatly exceeding the 1.7 million new electric vehicle registrations recorded in 2021 for the entirety of the European Union [57]. However, given the significant barriers to future development of hybrid deep geothermal projects including exploratory risks, financial uncertainty, and public opposition (see discussion), it is unlikely that 100% of the municipalities would be developed with combined geothermal–lithium plants. Nevertheless, if only 10% of the municipalities in the URG were to deploy such a plant, this could yield substantial benefits (see Figure 7). Assuming deployment would occur where geothermal potential is highest, total annual costs per municipality could be significantly reduced, with an average decrease of about 190%, whereas the total capacities of DHP and ORC plants would be about 655 $MW_{th}$ and 74 $MW_{el}$, respectively. More than 50 kt/a of lithium carbonate could be produced in these municipalities – enough to manufacture about 1.2 million electric vehicle battery packs annually.

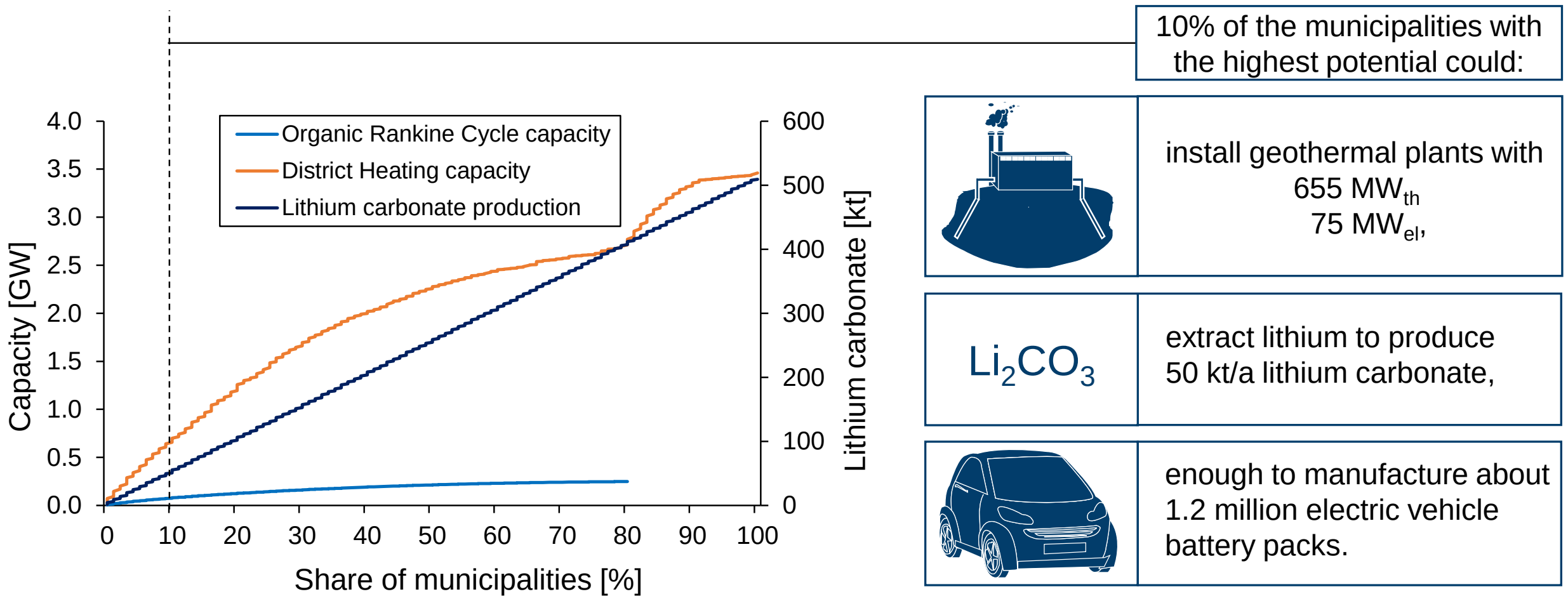

*Figure 7: Optimized capacities of Organic Rankine Cycles and district heating plants, as well as lithium carbonate production over the share of municipalities in the Upper Rhine Graben, whereas the share of 100% corresponds to 330. The municipalities are ordered by maximum achievable wellhead temperature (i.e., highest potential), as well as Organic Rankine Cycle capacity. The latter leads to the leap in the curve of district heating plant capacity, as no Organic Rankine Cycle plants are installed in the remaining municipalities.*

## 4. Discussion and Conclusions

Research on the extraction of lithium from geothermal brines dates to the early 1980s, while DLE technology has been in use for over 20 years at Livent Corporation's mine in Argentina [58]. Although the technology has been proven technically feasible with salar brines, uncertainties exist as to its application with geothermal brines, and its commercial efficacy remains to be proven. While presenting enormous potential, it is important to acknowledge that there has been a recent surge of hype with regard to geothermal lithium extraction that may exaggerate this potential [26]. One such example is that of Vulcan Energy's Zero Carbon Lithium project in the Upper Rhine Graben, which anticipates operating expenses roughly half those for geothermal–lithium operations in the Salton Sea area, despite having a significantly lower flow rate and lithium concentration [23,49]. Additional concerns regarding the sustainability of such lithium extraction are not without merit, as the geological source and refresh rate of these lithium deposits are not fully understood. Furthermore, social opposition, induced seismicity risks, and financial uncertainty could present major barriers to future development.

The geological source and refresh rate of lithium deposits are not yet fully understood; however, these factors may significantly impact results [29]. Geothermal brines are rich in minerals such as magnesium, potassium, and sodium and possess significant quantities of total dissolved solids, which may cause scaling in the geothermal plant, leading to the degradation of plant components and an increase in costs arising from maintenance and cleaning [25,59]. It is unknown how the addition of a lithium extraction facility would impact scaling and corrosion. In addition, the capital-intensive drilling phase is associated with considerable risk, which we accounted for in the model with conservative assumptions regarding the exploration costs. Subsurface geothermal resources are often not fully understood, and drilling may be unsuccessful in locating a hydrothermal resource with favorable characteristics for geothermal exploitation. Germany in general is considered a high-cost country for geothermal development, with drilling costs exceeding those in the U.S., for example. The risk of unsuccessful drilling can create significant financial losses and delays [60].

Furthermore, literature on deep geothermal energy, including the present article, focuses primarily on technical barriers to its use [61]. However, social acceptance is critical for the further deployment of geothermal plants. A seismic event attributed to a geothermal plant in Basel, Switzerland in 2006, with a magnitude reaching 3.4 on the Richter scale, marked a turning point in public perception of geothermal energy use in Germany and led to the emergence of a strong anti-geothermal protest movement [62]. Since then, incidents of subsidence and injection-induced seismicity with magnitudes of up to 2.6 in some German towns have solidified concerns about geothermal energy use [62,63]. The importance of social acceptance is illustrated in the example of the now-abandoned Brühl geothermal site in the Upper Rhine Graben, where construction of the planned geothermal plant was halted due a lack of public acceptance, despite drilling success and the achievement of high flow rates [64]. In addition to strategies for improving social acceptance, including preventing and minimizing undesirable effects, compensating local communities when damages occur, creating benefits for the latter, and enhancing community engagement [65], combined lithium extraction may also have a positive impact as "green lithium" and has received significant positive media coverage recently, and provides an attractive talking point for geothermal plant operators to present to the public.

If combined geothermal–lithium technology is not commercially-successful due to one of the above-mentioned reasons, the demand and environmental impacts of lithium procurement will potentially further increase. With current lithium supply insufficient to meet the anticipated 60-fold increase in lithium needed by 2050 to fulfill European Union demand, dependence on lithium imports from countries such as China, Australia, and Chile will likely increase, which could in turn impact the security of energy supply and transition to carbon-neutral energy systems. In addition, environmental and climate impacts associated with conventional lithium extraction will likely increase and lithium markets may become increasingly volatile due to highly concentrated supply [3]. If lithium market prices will also continue to rise, this could lead to new lithium resources being developed, especially carbon-intensive hard-rock deposits in Australia with a carbon footprint of about 15.8 kg $CO_{2,eq}$ per kg lithium carbonate equivalent [66]. This can be compared with estimated carbon footprints of 0.3 kg $CO_{2,eq}$ for brine deposits in South America [67]. Further research found that brine extraction has a carbon footprint of 3.2 kg $CO_{2,eq}$ and it is predicted that this will increase to 3.3 kg $CO_{2,eq}$ in 2100 [68]. The impacts are exacerbated by lithium having an estimated end-of-life recycling rate of less than 1% [69]. Assuming a carbon abatement potential of 15.8 kg $CO_{2,eq}$ when compared with conventional hard-rock procurement methods, the implementation of approximately 30 such geothermal-lithium plants in the Upper Rhine Graben could lead to an abatement of 800 kt $CO_2$ annually. Therefore, combined geothermal–lithium projects could present one of the best opportunities to decarbonize the lithium supply chain and could have a net negative carbon impact if the offsets of the generated power/heat are sold to the grid and displace coal-fired generation[64].

Given the numerous ongoing pilot projects demonstrating the potential of DLE from geothermal brines and the rapid advancement of the technology in recent years, the assumption of commercial success may be strengthened. With a total technical potential in Germany of 4155 $TWh_{el}$/a, deep geothermal energy could play a key role in the achievement of climate goals [70]. These geothermal plants could reduce $CO_2$ emissions from the energy sector and provide a much needed baseload supply of renewable heat and electricity not affected by weather and with a low land-use intensity [70]. The baseload heating is highly relevant in light of the energy crisis and desire to phase out imports of Russian natural gas [71]. Lithium extraction in combination with geothermal energy use could also increase and diversify lithium supply, reduce the environmental and climate impacts of lithium extraction, and aid in the energy transition by promoting the development of low-carbon technologies such as electric vehicle batteries and lithium-ion batteries for grid scale energy storage. Hybrid geothermal plants could also provide significant economic benefit in the form of stable jobs and a new domestic lithium industry in Germany, which possesses abundant lithium resources in the Upper Rhine Graben [5,29]. This lithium potential is not limited to Germany alone: significant lithium geothermal brine deposits have also been identified in the U.S., France, the U.K., and Italy [26,29] suggesting that the utilization of combined geothermal-lithium plants in future transformation strategies is essential.

**Data and Code Availability.** The ETHOS.FINE framework used is publicly available on GitHub (https://github.com/FZJ-IEK3-VSA/FINE). The TSAM tool for Pareto-optimal time series aggregation can also be found on GitHub (https://github.com/FZJ-IEK3-VSA/tsam). The potentials for renewable energies used in the optimizations are deposited on Zenodo (https://zenodo.org/record/6414018#.Y4m6bHbMI2w). The dataset for achievable hydrothermal temperatures in German municipalities is published together with a data article in the journal, Scientific Data (https://www.nature.com/articles/s41597-019-0233-0).

**Acknowledgments.** This work was supported by the German Federal Ministry for Economic Affairs and Climate Action (BMWK, 3EE5031D) and the Helmholtz Association as part of the program, “Energy System Design”. Due to the lack of published literature on this topic, many experts were consulted throughout the duration of this research. We would like to thank Prof. Ingrid Stober with the Department of Geology at the University of Freiburg, Dr. Thomas Kölbel and Elif Kaymakci of Energie Baden-Württemberg (EnBW) and the UnLimited project, Dr. Bernard Sanjuan of the French Geological Survey (BRGM), Dr. André Stechern of the Federal Institute for Geosciences and Natural Resources (BGR), and mineral commodities specialist Brian Jaskula of the U.S. Geological Survey (USGS) for their advice.

**CRediT-Statement.** Conceptualization: J.W.; data curation: G.V., J.W.; formal analysis: G.V., J.W.; funding acquisition: D.S.; investigation: G.V., J.W.; methodology: G.V., S.R., J.B., J.W.; software (FINE optimization model): S.R.; supervision: D.S.; validation: G.V., J.W.; visualization: G.V., J.W. ; writing – original draft: J.W., G.V.; writing – review and editing: S.R., J.B., N.P., J.L., D.S., G.V., and J.W.